\documentclass[aps,prl,twocolumn,superscriptaddress]{revtex4}

\usepackage{epsfig}
\usepackage{graphicx}
\usepackage{amsfonts}
\usepackage{amssymb}
\usepackage{amsmath}
\usepackage{bm}

\newcommand{\ckd}[1]{c^\dagger_#1}
\newcommand{\ck}[1]{c_#1}

\renewcommand{\vec}[1]{{\mathbf #1}}

\renewcommand{\vec}[1]{{\mathbf #1}}
\bibpunct{(}{)}{;}{s}{,}{,}

\makeatletter
\newcommand\appendix@section[1]{%
  \refstepcounter{section}%
  \orig@section*{Appendix \@Alph\c@section: #1}%
  \addcontentsline{toc}{section}{Appendix \@Alph\c@section: #1}%
}
\let\orig@section\section
\g@addto@macro\appendix{\let\section\appendix@section}
\makeatother

\begin{document}

\title{Topological Excitonic Superfluids in Three Dimensions}

\author{Youngseok Kim}
\affiliation{Department of Electrical and Computer Engineering, University of Illinois, Urbana, Il, 61801}
\altaffiliation{Micro and Nanotechnology Laboratory, University of Illinois, Urbana, Il 61801}

\author{E. M. Hankiewicz} 
\affiliation{Institut f\"{u}r Theoretische Physik und Astrophysik,
Universit\"{a}t W\"{u}rzburg, Am Hubland, 97074 W\"{u}rzburg, Germany} 

\author{Matthew J. Gilbert}
\affiliation{Department of Electrical and Computer Engineering, University of Illinois, Urbana, Il, 61801}
\altaffiliation{Micro and Nanotechnology Laboratory, University of Illinois, Urbana, Il 61801}

\begin{abstract}
We study the equilibrium and non-equilibrium properties of topological dipolar intersurface exciton condensates within time-reversal invariant topological insulators in three spatial dimensions without a magnetic field. We elucidate that, in order to correctly identify the proper pairing symmetry within the condensate order parameter, the full three-dimensional Hamiltonian must be considered. As a corollary, we demonstrate that only particles with similar chirality play a significant role in condensate formation. Furthermore, we find that the intersurface exciton condensation is not suppressed by the interconnection of surfaces in three-dimensional topological insulators as the intersurface polarizability vanishes in the condensed phase. This eliminates the surface current flow leaving only intersurface current flow through the bulk. We conclude by illustrating how the excitonic superfluidity may be identified through an examination of the terminal currents above and below the condensate critical current.
\end{abstract}
\pacs{71.35.-y, 73.20.-r, 73.22.Gk, 73.43.-f} 
\maketitle

Dipolar excitonic superfluidity (DES) has appeared in a veritable manifold of systems including microcavities~\cite{Balili:2007,Christopoulos:2007,Butov:2002}, cold atom systems~\cite{Baranov:2002,Neely:2010,Hadzibabic:2006,Potter:2010,Wang:2007} and semiconductor quantum wells~\cite{Kellogg:2004, Tutuc:2004, Snoke:2002, TiemannC:2008,Yoon:2010, Sinclair:2011}. Within condensed matter, emergent materials offer the possibility of finding new DESs. To this end, spatially segregated monolayers of graphene have been both theoretically\cite{Min:2008,Gilbert:2009,Zhang:2008} and experimentally\cite{Kim:2011} explored for signatures of excitonic superfluidity. While signs of interlayer correlation are experimentally observed, additional fermionic degrees of freedom, or flavors, screen the strength of the interlayer interaction\cite{Gilbert:2012} making the observation of DES in graphene multilayers challenging.    

The advent of time-reversal invariant topological insulators (TI) \cite{Hasan2010,QiRMP:2011} has brought renewed interest in finding DES in condensed matter systems. In sufficiently thin TI films, it has been proposed that spatially segregated surface electrons and holes may bind into a topological dipolar intersurface exciton superfluid (TDIES). To this point, existing approaches to TDIES have considered strictly two-dimensional Dirac surface states separated by an insulating spacer\cite{Seradjeh2009,Cho2011,Tilahun2011,Wang2011,Moon2012}. Yet the existence of a TDIES in three-dimensions is not a foregone conclusion based on two-dimensional surface state analysis. The most obvious drawback being that in a 3D TI, each of the surfaces is interconnected and there exists no obvious mechanism to segregate the electron and hole layers, as in other proposed systems.  

In this Letter, we theoretically study the equilibrium and non-equilibrium properties of TDIES in 3D TI and show that a stable TDIES may be formed. We link this stability of TDIES  in 3D TI  to the fact that intersurface polarizability vanishes in the TDIES phase forbidding quasiparticle recombination via single particle mechanisms.  Further, we find that in order to obtain the proper form of the condensate order parameter, the full 3D Hamiltonian must be used. We propose that the TDIES phase may be observed via examination of the terminal currents via 4-terminal electrical transport measurements.  
\begin{figure}[t!] %FIGURE1=============================
\centering
\includegraphics[width=0.5\textwidth]{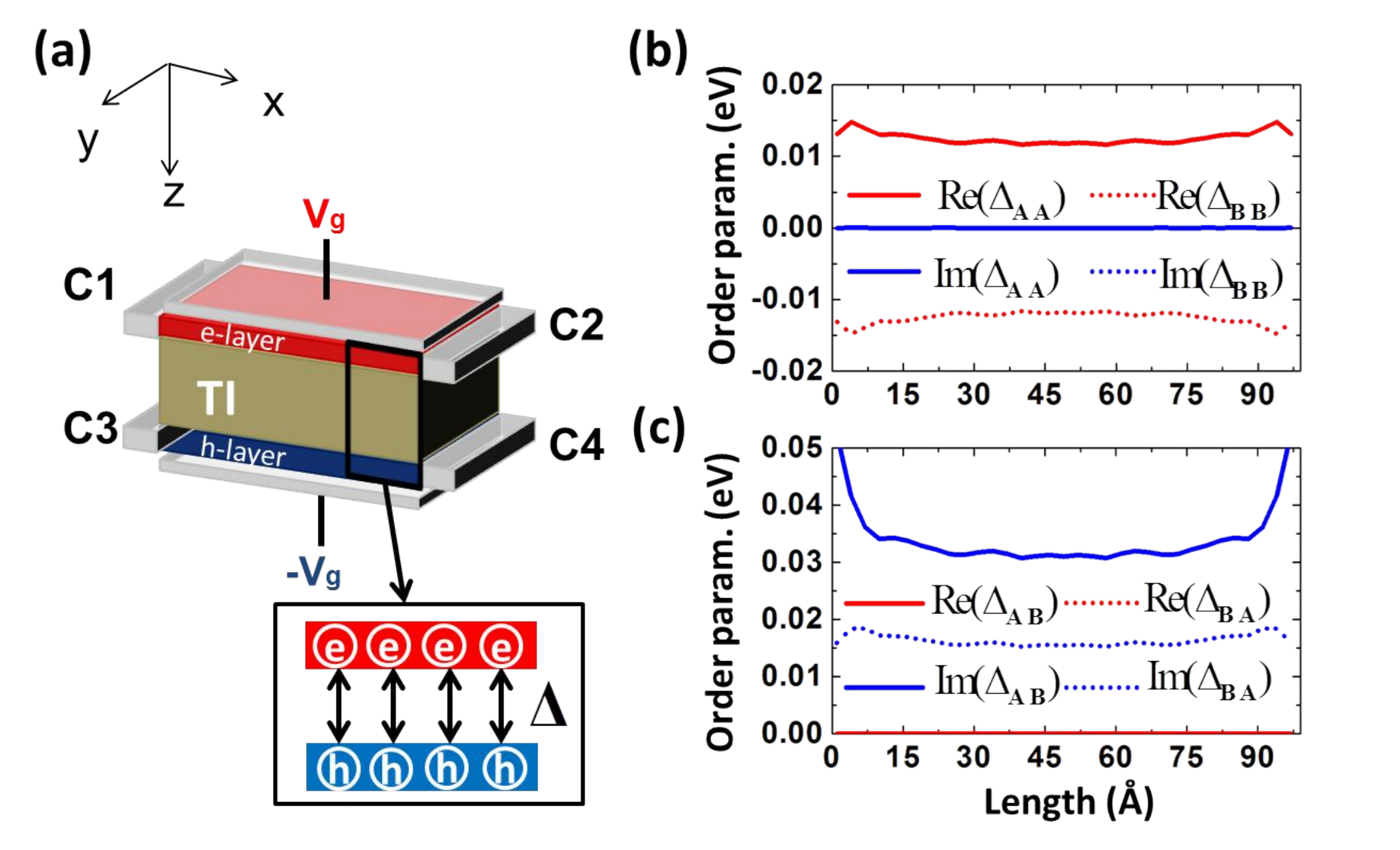}
\caption{(a): Schematic of topological insulator thin-film system under consideration. The top and bottom surfaces are assumed to contain equal numbers of electron and holes, respectively.
 (b, c): A plot of order parameter (b) $U\mathbf{\Delta}_{\tau,\sigma}$ and (c) $U\mathbf{\Delta}_{\tau\prime,\sigma}$ as a function of device length at the middle of the width. Two matrix elements of $\Delta_{A\uparrow,A\uparrow}$ (solid line) and $\Delta_{B\uparrow,B\uparrow}$ (broken line) are plotted in (b) and two matrix elements of $\Delta_{A\uparrow,B\uparrow}$ (solid line) and $\Delta_{B\uparrow,A\uparrow}$ (broken line) is plotted in (c). The real and imaginary parts are plotted in red and blue colors, respectively, and $U=1.5~{\rm eV}$.}
\label{fig1}
\end{figure} %FIGURE1=================================

% system modeling part start ================ 
We begin in Fig. \ref{fig1}(a), where we schematically show the system we consider. We apply top and bottom potentials of opposite polarity to induce electrons on the top surface and holes on the bottom surface. We attach contacts C1 - C4 to the top and bottom surfaces on the left and right sides of the TI through which current may be injected and extracted as seen in Fig. \ref{fig1}(a). The Hamiltonian for our system is that of a 3D time-reversal invariant TI\cite{ Liu2010, Chiu2011}
%single particle Hamiltonian
\begin{equation} \label{eq:H}
{\mathbf H}_0=\sum_{\mathbf k}\ckd{k}[d_a({\mathbf k})\Gamma^a+M({\mathbf k})\Gamma^0]\ck{k},
\end{equation}
with $a=x,y,z$. In Eq. (\ref{eq:H}), our basis includes two orbital components ($A,B$ corresponding for example to $P1^{+}_{-}$ and $P2^{-}_{+}$  in Bi$_2$Se$_3$ \cite{Liu2010}) and the two spin components ($\uparrow,\downarrow$). Within our basis, the annihilation operator is defined as $\ck{k}=( c_{k,A\uparrow} c_{k,B\uparrow} c_{k,A\downarrow} c_{k,B\downarrow} )$. We define the requisite gamma matrices in Eq. (\ref{eq:H}) as $\Gamma^a=\sigma^a\otimes\tau^x$, $\Gamma^0={\rm {\mathbf I}}\otimes\tau^z$ where $\tau^a$ and $\sigma^a$ are Pauli matrices for orbital and spin, respectively. Additionally, in Eq. (\ref{eq:H}), $d_a({\mathbf k})=\hbar v_F k_a$ and $M({\mathbf k})=m-(1/2)bk^2$ where $v_F$, $m$, and $b$ are materials dependent parameters.
In Eq. (\ref{eq:H}), the topological states occur when $m/b>0$.\cite{Chiu2011} In this work, we set $\hbar v_F=3~eV\AA$, and $b=-9~eV\AA^2$ with an applied surface gate bias of $V_g=1.0~V$ which places us in the dense electron-hole regime where we expect the pairs to form a BCS-type state. In our model the value of $m$ is set to $-1.5~eV$. This value of $m$ ensures that the surface states are localized in $\hat{z}$-direction within one lattice constant.\footnote{We select a large $m$ solely to ease the computational burden. The results presented here would not change qualitatively when we use the material parameters for Bi$_2$Se$_3$ instead of the model Hamiltonian parameters.} 
The full single particle Hamiltonian is then Fourier transformed into the real space assuming low energy excitations to obtain the single particle lattice Hamiltonian (see Supplementary A) where the lattice constant is set to be $a_0=3~\AA$.

%interaction Hamiltonian

With the non-interacting Hamiltonian defined, we now specify the intersurface interactions. As long as the chemical potential remains within the bulk gap, the surface state wavefunctions decay exponentially as a function of distance from the surface. As such, we may define the interactions purely as 2D intersurface interactions between the top and bottom surface through a local density approximation, ${\mathbf H}_{\rm int} =-\sum_{<i,j>}U_{i,j}n(i)n(j)$, with $n(i)=\sum_s c_s^\dagger(i)c_s(i)$ being the electron density operators at a lattice site $i$ with spin and orbital index of $s=A\uparrow, B\uparrow, A\downarrow, B\downarrow$. We assume an attractive intersurface interaction mediated by Coulomb interactions, $U_{i,j}=U\delta_{i,j}$, as such we simplify the intersurface interaction Hamiltonian as,
\begin{equation} \label{Eq:Hint}
{\mathbf H}_{\rm int} = -U\sum_{i}\sum_{s,s'}e^\dagger_{s}(i)e_{s}(i)  h^\dagger_{s'}(i)h_{s'}(i).
\end{equation}
In Eq. (\ref{Eq:Hint}), we define annihilation operator of top surface (electron layer) as $e(i)$ and bottom surface (hole layer) as $h(i)$ at an in-plane lattice site $i$. Following standard mean field decomposition, we may finally obtain our intersurface interaction contribution as
\begin{equation} \label{Eq:Hint2}
\begin{split}
\mathbf{H}_{\rm int}&\simeq U\sum_{i}\sum_{s,s'} \left[ \Delta_{s,s'}(i) h^\dagger_{s'}(i)e_{s}(i) \right. \\
&\left. + \Delta^\dagger_{s,s'}(i) e^\dagger_{s}(i)h_{s'}(i) - \vert \Delta_{s,s'}(i) \vert^2 \right]. \\
\end{split}
\end{equation}
We define the exciton order parameter as\cite{su:2008, Shim2009}
\begin{equation} \label{Eq:delta} 
\Delta_{s,s'}(i)=\langle e^\dagger_{s}(i)h_{s'}(i) \rangle,
\end{equation}
%
%Begin Order Parameter Discussion
With the order parameter phase expressed as
\begin{equation} \label{Eq:delta2} 
\begin{split}
%\vert \Delta_{s,s'} \vert=\sqrt{(\Delta^x_{s,s'})^2+(\Delta^y_{s,s'})^2},\\
\varphi_{s,s'}=\tan^{-1}\left( \Delta^y_{s,s'}/\Delta^x_{s,s'}\right),\\
\end{split}
\end{equation}
where the $\Delta^x_{s,s'}$ and $\Delta^y_{s,s'}$ are real and imaginary parts of the order parameter $\Delta_{s,s'}$. 

Using the total Hamiltonian of $\mathbf{H}=\mathbf{H}_0+\mathbf{H}_{\rm int}$, we may study the equilibrium properties of the system. We turn our focus to the TDIES order parameter which is obtained by diagonalizing the system Hamiltonian $\mathbf{H}$ with the system temperature, $T_{sys}=0~K$ (see Supplementary B). As our Hamiltonian has both orbital and spin degrees of freedom, there are four possible pairings in the order parameter described in Eq. (\ref{Eq:delta}). To clarify this point, we define exciton order parameter subset as $\mathbf{\Delta}_{\tau,\sigma},~\mathbf{\Delta}_{\tau',\sigma},~\mathbf{\Delta}_{\tau,\sigma'}$, and $\mathbf{\Delta}_{\tau',\sigma'}$ where $\tau~(\sigma)$ stands for the same orbital (spin) pairing, while $\tau'~(\sigma')$ stands for different orbital (spin) pairing (e.g. $\mathbf{\Delta}_{\tau,\sigma}\supset\lbrace \Delta_{A\uparrow,A\uparrow}~\Delta_{B\uparrow,B\uparrow}~\Delta_{A\downarrow,A\downarrow}~\Delta_{B\downarrow,B\downarrow}\rbrace$). We calculate the order parameter self-consistently in a structure of dimensions $99(\hat{x})\times33(\hat{y})\times24(\hat{z})~\AA$, and recognize that only two types of the exciton pairing order parameters are non-zero: pairing between the same spin and the same orbital ($\mathbf{\Delta}_{\tau,\sigma}$) shown in Fig. \ref{fig1}(b) and pairing between the same spin but different orbitals ($\mathbf{\Delta}_{\tau',\sigma}$) shown in Fig. \ref{fig1}(c). $\mathbf{\Delta}_{\tau',\sigma}$ is of particular importance as it has not been described in the previous work involving the effective single surface model\cite{Seradjeh2009}. We see that $\mathbf{\Delta}_{\tau,\sigma}$ is purely real while $\mathbf{\Delta}_{\tau',\sigma}$ is purely imaginary. To understand which of these is correct, we calculate the ground state energy of system, which is minimized when we choose $\mathbf{\Delta}_{\tau',\sigma}$ (see Supplementary C). The argument of the intersurface phase relationship is also consistent with the result. $\mathbf{\Delta}_{\tau',\sigma}$ has  $\varphi_{s,s'} = \pi/2$ whereas $\mathbf{\Delta}_{\tau,\sigma}$ has $\varphi_{s,s'} = 0$. Therefore, the purely imaginary order parameter is proper as it corresponds to the intersurface phase relationship which maximizes intersurface coherence. With $\varphi_{s,s'} = 0$ no TDIES exists as the surfaces are completely decoupled. Closer examination of the order parameter reveals a dependence on quasiparticle chirality, in which only electrons and holes with identical chirality bind. 
\begin{figure}[t!] %FIGURE2=============================
\centering
\includegraphics[width=0.5\textwidth]{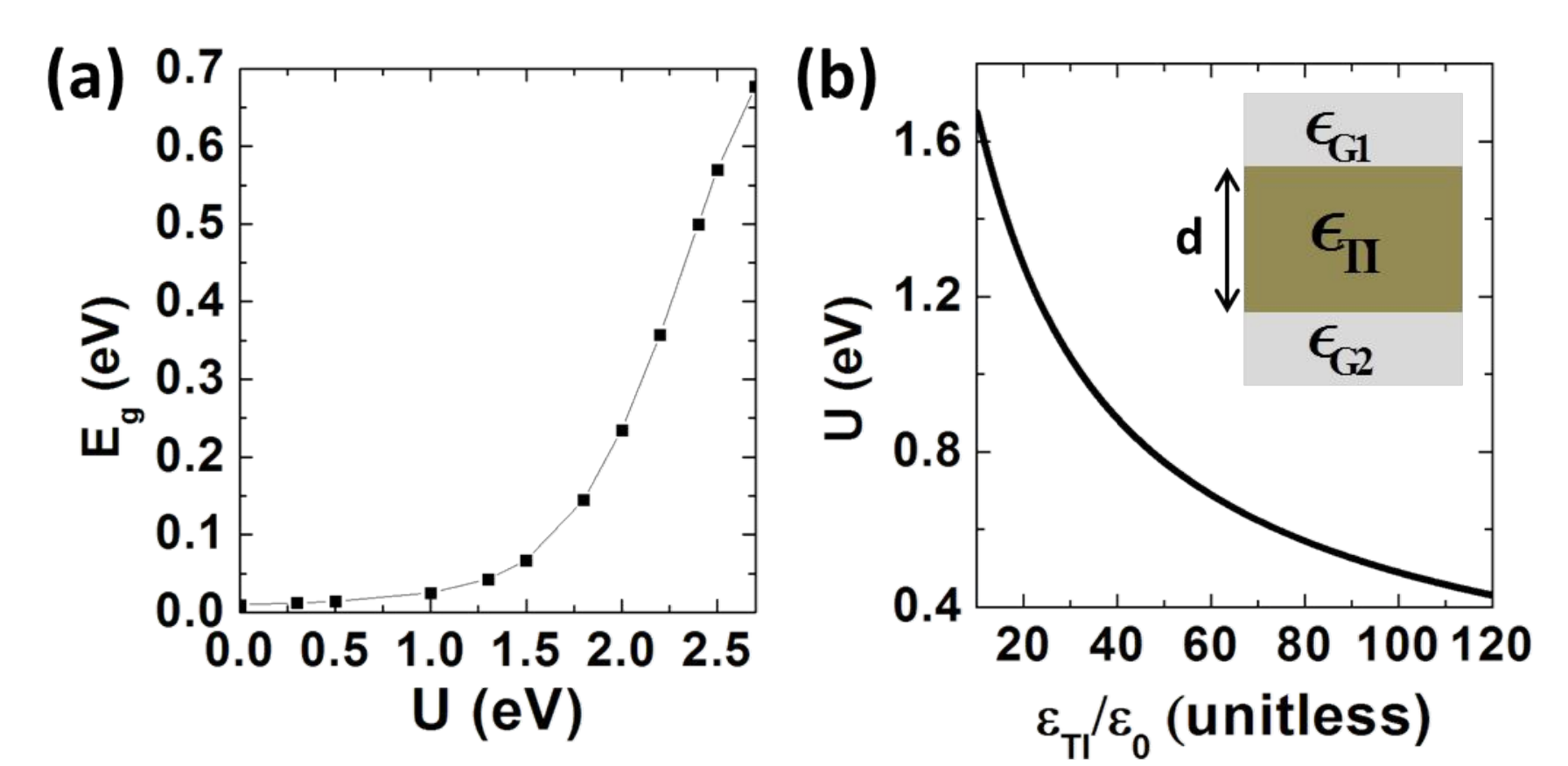}
\caption{(a): A plot of the intersurface interaction induced energy gap as a function of interaction constant, $U$ with the gate bias $V_g=1.0~V$. The non-zero $E_g$ for $U=0~ eV$ originates from the finite-size effects. (b): A plot of intersurface interaction strength, $U$, as a function of dielectric constant of topological insulator, $\epsilon_{TI}$. The thickness of the TI is fixed at $d=24~\AA$. The inset illustrates a schematic dielectric structure. }\label{fig2}
\end{figure} %FIGURE2=================================
%
% End Order Parameter Discussion
% Start a Discussion About Figure 2 ====================

Beyond the pairing symmetry, we must understand the size of the interaction induced gap in a TDIES. Fig. \ref{fig2}(a) shows the size of the self-consistent interaction induced gap as a function of the interaction strength, $U$. Yet we know that the intersurface interaction is influenced by the dielectric environment. To understand this effect, we consider the bulk dielectric constant of a TI ($\epsilon_{TI}$) which is in contact with top and bottom surface insulating layers having dielectric constants of $\epsilon_{G1}$ and $\epsilon_{G2}$, respectively (see inset of Fig. \ref{fig2}(b)). In this case, the bare intersurface Coulomb interaction is given by\cite{Profumo2010,Tilahun2011}:
\begin{equation}
\label{Eq:screeningq} 
\tilde{U}_{tb}(q) = \frac{8 \pi e^2}{q D(q) } \;  \epsilon_{TI},
\end{equation}
where $D(q) =  (\epsilon_{G1}+\epsilon_{TI})  (\epsilon_{TI}+\epsilon_{G2})\, e^{qd}+(\epsilon_{G1}-\epsilon_{TI})  (\epsilon_{TI}-\epsilon_{G2})\, e^{-qd}$ and $d$ is the intersurface separation and $q$ is the wavevector. 
From this, it is possible to estimate interaction strength in real space, $U_{tb}(r)$, where $r$ is the planar radius, using the Fourier transformation of $\tilde{U}_{tb}(q)$. We are particularly interested in the case of $r=0$, as Eq. (\ref{Eq:Hint}) only considers local intersurface interactions. Using a TI thickness of $d=24~\AA$, we obtain the resultant intersurface interaction strength $U_{tb}(0)=U$ as a function of the TI dielectric constant, as illustrated in Fig. \ref{fig2}(b). In this work, we select an intersurface interaction strength of $U=1.5~eV$ in order to ensure a large enough gap ($E_g\simeq67$ meV), that we may observe distinct characteristics of the condensate phase at a finite-size system while not unrealistically large in magnitude. By taking the thin-film limit, $qd \to 0$, we find that $\tilde{U}_{tb}(q)$ goes to $4 \pi e^2/q(\epsilon_{G1}+\epsilon_{G2})$, and is independent of $\epsilon_{TI}$. For this reason, although the dielectric constant of real material is large and results in a reduced intersurface interaction, the thin film limit ensures a considerable intersurface interaction strength. Even in a thin film limit, however, the material should have a low level of bulk doping, since the intersurface interaction is effectively screened by a doped bulk (see Supplementary D).
 
%End the Discussion about Figure 2=======================
With an understanding of the equilibrium properties, we now seek to understand the salient non-equilibrium properties through the application of the non-equilibrium Green's function formalism\cite{Datta:2000}. We use a structure of size $195(\hat{x})\times33(\hat{y})\times24(\hat{z})~\AA$ in transport calculations to ensure sufficient lateral separation of the surface contacts during current injection. We iterate over the Green's function and the intersurface interactions until the $\Delta_{s,s'}$ reaches self-consistency. Once the self-consistency is achieved, the contact and spatially resolved currents (see Supplementary E) are calculated. 

One of the key questions concerning the utimate stability of the TDIES in 3D arises from the nature of a 3D TI. In a 3D TI, each of the surfaces is interconnected and the single-particle hopping term may easily compete with the many-body intersurface interaction. Therefore, it may be more energetically favorable for electrons on one surface to annihilate holes on the other surface via the adjoining surface rather than forming a TDIES. As such, elucidating where the current flows in our system is one of the most crucial questions. To drive current flow, we choose the drag-counterflow bias configuration in which $V_1=-V_{bias},~V_2=V_{bias}$ and $V_3=V_4=0~V.$\cite{su:2008} This configuration will drive an intersurface current flow from the top surface to bottom surface on the left side of our system and from the bottom surface to the top surface on the right hand side. When the system is in a TDIES phase, then we expect this to be the only mechanism for current flow with the superfluid gap forbidding transport across either the electron or hole doped surface. 
\begin{figure}[t!] %FIGURE3=============================
\centering
\includegraphics[width=0.5\textwidth]{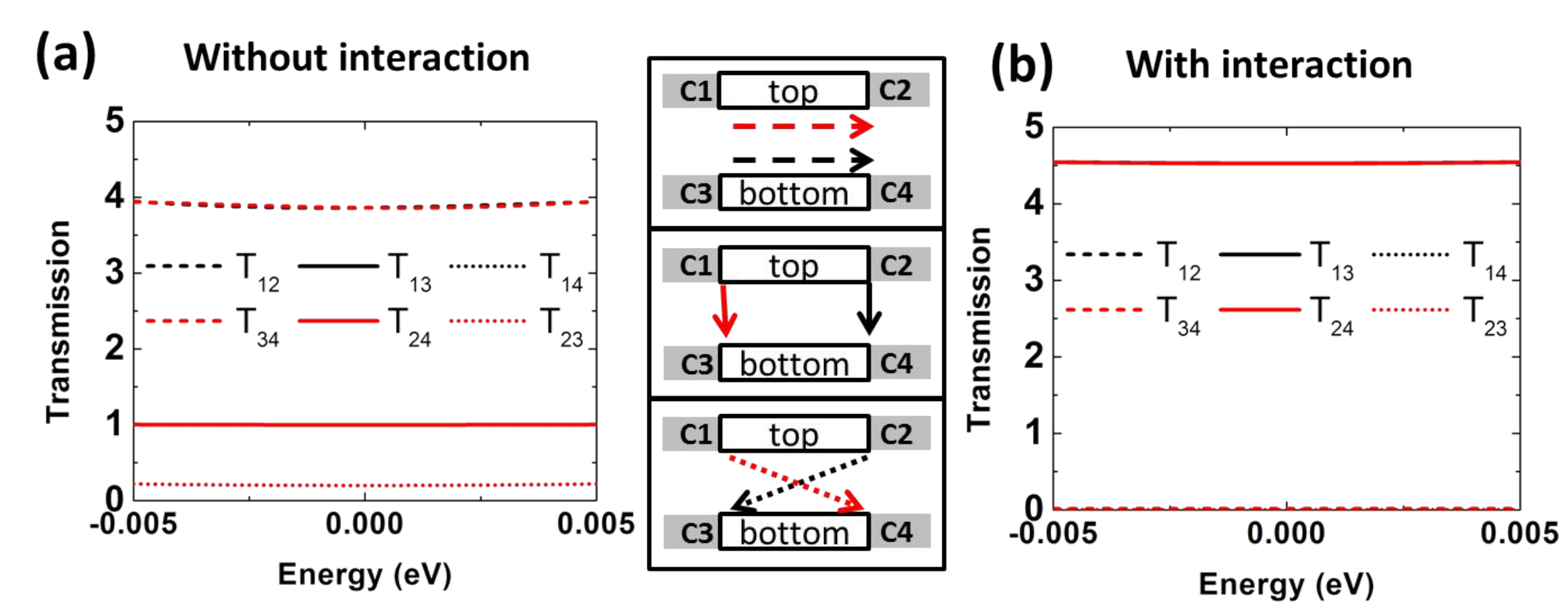}
\caption{A transmission between C1, C2, C3, and C4 of the system (a) without and (b) with intersurface interaction at a bias of $V_{bias}=5~mV$. (a): When there is no interaction and the system is not gapped, current flows across the device resulting in $T_{12}$ (or $T_{34}$) dominating the transport. (b): When the system is gapped, however, the quasiparticle undergoes a similar process to Andreev reflections, resulting in large $T_{13}$ and $T_{24}$ as current flows from top surface to bottom surface with no transmission across either of the top $T_{12}$ or bottom surfaces $T_{34}$. }\label{fig3}
\end{figure} %FIGURE3=================================

In Fig. \ref{fig3}, we plot the resultant transmissions from each contact to every other contact at an intersurface bias which drives a current well below that of the superfluid critical current without and with intersurface interactions. Here we choose to deal with transmissions to track the quasiparticle motion. Normally, one would simply examine the terminal currents, however, with the surfaces interconnected, there is no way to determine the current path to a particular contact. Furthermore, as we are well within the linear response regime, the examination of individual transmissions will not substantially differ within the energy integral and remain a valid method to assess current flow. When there are no intersurface interactions, as in Fig. \ref{fig3}(a), the transport properties are dominated by transmissions directly across the surfaces ($T_{12}$ and $T_{34}$), however, direct transmissions from the top surface to the bottom surface ($T_{13}$ and $T_{24}$) and diagonal intersurface transmissions ($T_{14}$ and $T_{23}$) are non-negligible. This is understood by noting that, although we are driving a current across the top surface, the presence of gapless states on each of the interconnected surfaces contributes to the total contact current. 

This is to be contrasted with Fig. \ref{fig3}(b) where the intersurface interactions are included. In this case, direct intersurface transmissions dominate the transport characteristics while transport both across individual surfaces and diagonal intersurface transport are negligible. This signals the acquisition of a gap corresponding to the formation of a TDIES. Additionally, we find no diagonal intersurface transport yet the side surface connecting the top and bottom layer is not gapped. The lack of surface current flow between the top and bottom surfaces lies in the fact that when the TDIES is formed, each of the constituent electrons and holes is paired. This forces the intersurface polarizability to drop to zero as there are no free charges available on either the top or bottom surfaces to respond to voltage perturbations\cite{Min:2008}. As we are in the dense electron-hole regime, we expect the intrasurface polarizability to be zero before the onset of TDIES. More mathematically, the static intersurface polarization operator\cite{Bistritzer08} $\Pi = g \sum_k  \partial_{E_k} n_F(E_k)$ , where g = 2  is the number of fermionic degrees of freedom in our system, and $E_k$ is the energy of a quasiparticle in the condensed state. Since the condensate acquires a gap, $\Pi$  must vanish at zero temperature and this is a critical insight into the formation of a stable TDIES without the necessity of gapping the side surfaces to force intersurface segregation. 
%
% Fig. 4 starts ======================
%
\begin{figure}[t!] %FIGURE4=============================
\centering
\includegraphics[width=0.5\textwidth]{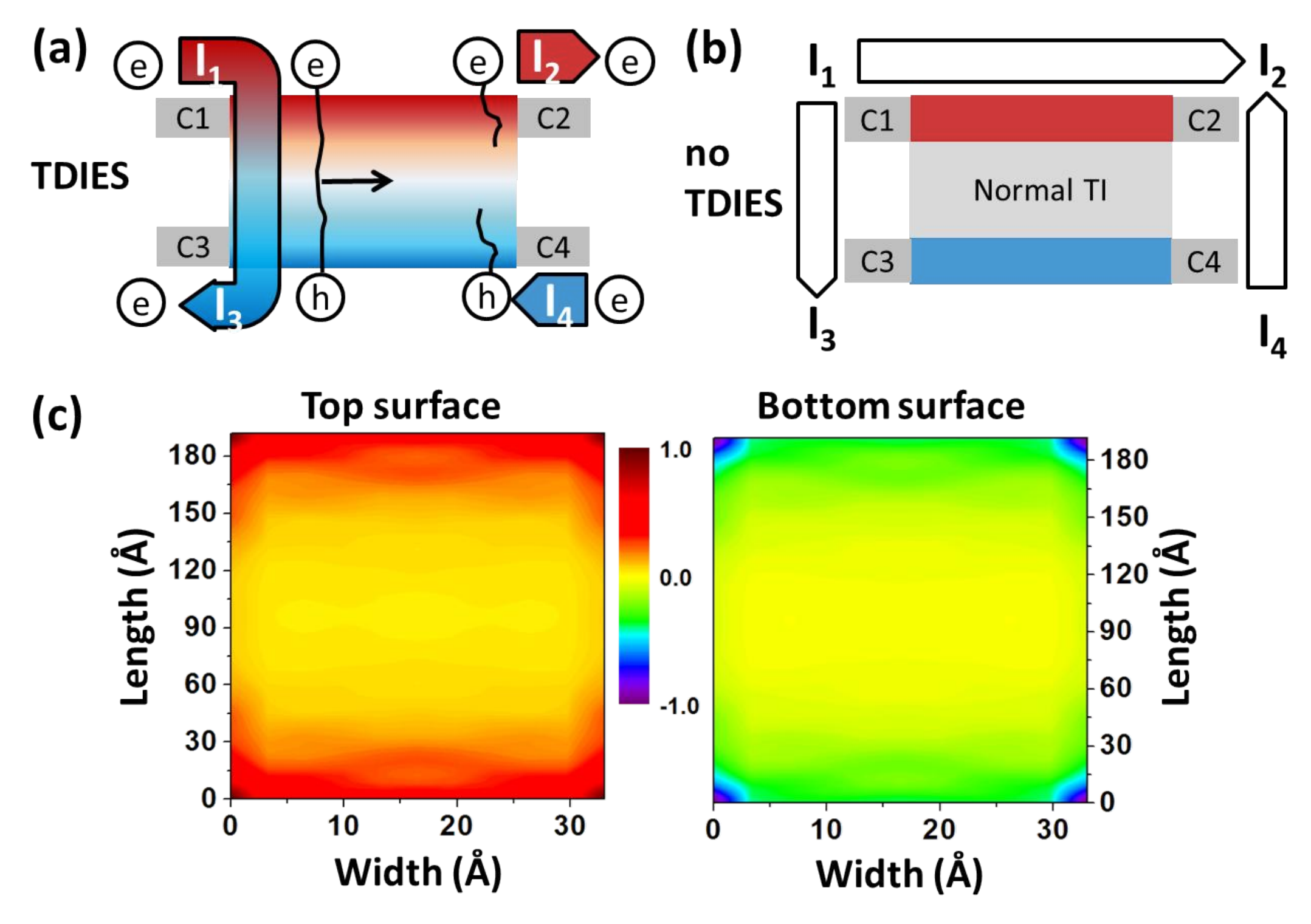}
\caption{A schematic of the transport directed ($\hat{x}$) current flow (a) with and (b) without intersurface interaction. (a): In the presence of TDIES, current flows between C1 and C3 on the left and from C2 to C4 on the right. (b): However, in the system without gap, all surfaces are interconnected and the current flows mainly from C1 to C2.
(c): A plot of spatially resolved transport directed current at the top (left) and bottom (right) surfaces with $V_{bias}=5~mV$. The quasiparticle tunneling process within the coherence length is manifested as an equal amounts of current with a different sign on opposite surfaces. The current is normalized by $|I_{max}|=0.11~{\rm \mu A}$.}\label{fig4}
\end{figure} %FIGURE4=================================

In as much as the individual transmissions provide important insight into where the currents flow, directional current densities provide additional clarity. The process for the non-equilibrium conduction is shown in Fig. \ref{fig4}(a). When an electron is injected into the top layer, via C1, within the energy range of the superfluid gap, the electron undergoes coherent transport within a characteristic distance away from the injecting contact, the coherence length, $L_c$. Beyond $L_c$, the injected electron is retroreflected to the opposite paired surface with opposite momentum, leading to a significant intersurface current flow, and exiting the system through C3. To conserve current, the system launches the exciton across the system which breaks when the exciton reaches the contacts on the opposite side of the surfaces\cite{su:2008,Gilbert2010}. This is not the case in Fig. \ref{fig4}(b) where we do not have a TDIES in the system. The top and bottom surfaces are not gapped and this allows transfer of charge across the top surface from C1 to C2. However, as none of the other surfaces are gapped we also expect to see charge transfer from the bottom contacts reach the top surface contact (e.g. from C4 to C2). 
In the Fig. \ref{fig4}(c), we show the spatially resolved current in the transport, or $\hat{x}$, direction in top and bottom surfaces with TDIES. In this scenario,  we exactly see the physical manifestation of the exciton flow described in Fig. \ref{fig4}(a) as an identical amount of current flows in top and bottom surfaces with different direction within $L_c \approx 3~nm$ away from the contacts. This fact, allows for a simple electrical measurement to detect the presence of a TDIES.  When the system is biased in the drag counterflow configuration and the system remains below the superfluid critical current, the amount of current transferred between the two surfaces at each respective side of the system is identical ($I_{1} = - I_{3}$ and $-I_{2} =  I_{4}$). However, above critical current, in condensate gap closes and the terminal currents are no longer equal and opposite on the respective sides as single particle processes dominate the surface and intersurface transport\cite{su:2008}. 

In conclusion, we have examined TDIES in 3D. We find that the proper exciton pairing order parameter is purely imaginary which is necessary to properly account for the system dynamics. Furthermore, we find that the exciton order parameter is p-wave and that electrons will only bind with holes of the same chirality from different orbitals. We find that TDIES in time-reversal invariant TI thin films prefer to bind into excitons with a dominant many-body energy which prevents single particle intersurface transport as the intersurface polarizablity drops to zero. This allows for the observation of superfluid behavior out of the quantum Hall regime and without the need to artificially segregate the surfaces. Finally, we find that the presence of a topological superfluid may be electrically detected by the presence of equal and opposite intersurface contact currents.

\begin{acknowledgements}
YK acknowledges insightful discussions with P. Ghaemi, H. -H. Hung and X. Chen. MJG thanks T. L. Hughes for useful discussions. We acknowledge support for the Center for Scientific Computing from the CNSI, MRL. YK is supported by Fulbright Science and Technology Award. EMH acknowledges funding through DFG Grant HA 5893/3-1. MJG acknowledges support from the Army Research Office (ARO) under contract number W911NF-09-1-0347, the Office of Naval Research (ONR) under contract number N0014-11-1-0728, and the Air Force Office of Scientific Research (AFOSR) under contract number FA9550-10-1-0459.
\end{acknowledgements} 
%%%%%%%%%%%%%%%%%%%%%%%%%%%%%%%%%%%%%%%%%%%%%%%
%%%%%%%%%%%%%%%%%%%%%%%%%%%%%%%%%%%%%%%%%%%%%%%
%Appendix
\appendix
\section{Lattice Model Hamiltonian for Time-Reversal Invariant Topological Insulator} \label{A:H}
Within a lattice model description, assuming a low energy excitation, the $d_a({\mathbf k})=\hbar v_F k_a$ and $M({\mathbf k})=m-(1/2)bk^2$ in Eq. (1) is read as
\begin{equation} \label{Eq:dM}
\begin{split}
&d_a({\mathbf p})=(\hbar v_F/a_0) \sin(p_a a_0),\\
&M({\mathbf p})=m-3b/a_0^2 \\
&~~~~~~~~+ (\cos(p_x a_0) +\cos(p_y a_0) + \cos(p_z a_0) ),\\
\end{split}
\end{equation}
for a given lattice constant $a_0$. As a result, Eq. (1) will be described as the following form in a lattice model:
\begin{equation}
\begin{split} \label{Eq:H2}
{\mathbf H}_0=&\sum_{\mathbf k}\ckd{k} \left[  \left( m-\frac{3b}{a_0^2} \right)\Gamma^0 \right.\\
& \left. + \sum_a \left( \frac{\hbar v_F}{a_0} \sin(k_a a_0) \Gamma^a+\frac{b}{a_0^2}\cos(k_a a_0) \Gamma^0  \right)   \right] \ck{k}.\\
\end{split}
\end{equation}
The sine and cosine terms in Eq. (\ref{Eq:H2}) are Fourier transformed into nearest neighbor terms in the real space Hamiltonian and, as a result, we obtain
\begin{equation} \label{Eq:H3}
\begin{split}
{\mathbf H}_0=&\sum_{xyz}c^\dagger_{xyz} \left[ \left( m-\frac{3b}{a_0^2} \right) \Gamma^0 \right] c_{xyz} + \\
&c^\dagger_{xyz} \left[ \frac{b}{2a_0^2}\Gamma^0 - \frac{i\hbar v_F}{2a_0}\Gamma^x \right] c_{x+1yz} + H.C.\\%c^\dagger_{xyz} \left[ \frac{b}{2a_0^2}\Gamma^0 + \frac{i\hbar v_F}{2a_0}\Gamma^x \right] c_{i-1jk} + \\
&c^\dagger_{xyz} \left[ \frac{b}{2a_0^2}\Gamma^0 - \frac{i\hbar v_F}{2a_0}\Gamma^y \right] c_{xy+1z} + H.C.\\%c^\dagger_{xyz} \left[ \frac{b}{2a_0^2}\Gamma^0 + \frac{i\hbar v_F}{2a_0}\Gamma^y \right] c_{ij-1k} + \\
&c^\dagger_{xyz} \left[ \frac{b}{2a_0^2}\Gamma^0 - \frac{i\hbar v_F}{2a_0}\Gamma^z \right] c_{xyz+1} + H.C.\\%c^\dagger_{xyz} \left[ \frac{b}{2a_0^2}\Gamma^0 + \frac{i\hbar v_F}{2a_0}\Gamma^z \right] c_{xyz-1}. \\
\end{split}
\end{equation}
%%%%%%%%%%%%%%%%%%%%%%%%%%%%%%%%%%%%%%%%%%%%%%%
\section{Order Parameter Calculation} \label{A:delta}
It is possible to calculate order parameter $\Delta_{s,s'}$ in Eq. (4) via diagonalization of the system Hamiltonian ${\mathbf H}={\mathbf H}_0+{\mathbf H}_{\rm int}$. Assuming ${\mathbf H}$ is hermitian, the corresponding eigenvalue (${\mathbf D}$) and eigenvector (${\mathbf V}$) matrix satisfy following relationship,
\begin{equation} \label{Eq:VDV}
\begin{split}
&{\mathbf H}={\mathbf V}{\mathbf D}{\mathbf V}^{-1}={\mathbf V}{\mathbf D}{\mathbf V}^{\dagger} \\
&=
\left( 
\begin{matrix} 
\frown & \frown & \frown & \frown\\
&&&\\
{\mathbf v_1} & {\mathbf v_2} & \cdots & {\mathbf v_N}\\
&&&\\
\smile&\smile&\smile&\smile\\
\end{matrix}\right) \left( 
\begin{matrix}
\varepsilon_1 & 0 & \cdots & 0\\
0 & \varepsilon_2 & \cdots & 0\\
\vdots & \vdots & \ddots & \vdots\\
0 & 0 & \cdots & \varepsilon_n\\
\end{matrix}\right) \left(
\begin{matrix}
(&& {\mathbf v_1}^\dagger &&)\\ 
(&& {\mathbf v_2}^\dagger &&)\\ 
&& \vdots &&\\ 
(&& {\mathbf v_N}^\dagger &&)\\
\end{matrix}\right) \\
\end{split}
\end{equation}
where the column vector ${\mathbf v_m}$ corresponds to a normalized eigenvector of the eigenvalue of $\varepsilon_m$ with a total number of eigenvalues N. As a result, the eigenstate $\gamma_m$ with corresponding energy $\varepsilon_m$ is connected to the states at a lattice site $l$ ($c_l$) via a mapping rule of $\mathbf{V}$ as follow,
 \begin{equation} \label{Eq:mapping}
\begin{split}
& \left[{\mathbf \gamma}^\dagger \right]= \left( 
\begin{matrix} 
{\mathbf \gamma_1}^\dagger\\ 
{\mathbf \gamma_2}^\dagger\\ 
 \vdots \\ 
{\mathbf \gamma_N}^\dagger\\
\end{matrix}\right) = \left(
\begin{matrix}
(&& {\mathbf v_1}^\dagger &&)\\ 
(&& {\mathbf v_2}^\dagger &&)\\ 
&& \vdots &&\\ 
(&& {\mathbf v_N}^\dagger &&)\\
\end{matrix}\right) \left(
\begin{matrix}
{\mathbf c_1}^\dagger\\ 
{\mathbf c_2}^\dagger\\ 
 \vdots \\ 
{\mathbf c_N}^\dagger\\
\end{matrix}\right) 
={\mathbf V}^\dagger \left[{\mathbf c}^\dagger \right]. \\
\end{split}
\end{equation}
Using the matrix identity of ${\mathbf V}{\mathbf V}^\dagger={\mathbf V}^\dagger{\mathbf V}=\mathbb{I}$, it is easy to map eigenstates to real space basis as
$\left[{\mathbf c}^\dagger \right]={\mathbf V}\left[{\mathbf \gamma}^\dagger \right]$. whose matrix element and its complex conjugate are
\begin{equation}\label{Eq:c-gamma}
c_l^\dagger=\sum_m^N V_{lm}\gamma_m^\dagger,~~c_l=\sum_m^N V_{lm}^\dagger\gamma_m,
\end{equation}
where $V_{lm}$ stands for a $l,m$ component matrix element of ${\mathbf V}$. As a result, the order parameter in Eq. (4) can be calculated as
\begin{equation}\label{Eq:delta3}
\begin{split}
\Delta(l,l')&=\langle c_l^\dagger c_{l'} \rangle 
=\left\langle \left( \sum_m^N V_{lm}\gamma_m^\dagger \right) \left( \sum_{m'}^N V_{l'm'}^\dagger\gamma_{m'} \right) \right\rangle \\
&=\sum_{m,m'}^N V_{lm}V_{l'm'}^\dagger \langle \gamma_m^\dagger\gamma_{m'} \rangle \\
%&=\sum_{m,m'}^N V_{lm}V_{l'm'}^\dagger \delta_{mm'} f(\varepsilon_m-\mu) \\
&=\sum_{m}^N V_{lm}V_{l'm}^\dagger f(\varepsilon_m-\mu) , \\
\end{split}
\end{equation} 
where $f$ is the Fermi-Dirac distribution and $\mu$ is the bulk chemical potential. The first line of the Eq. (\ref{Eq:delta3}) is from Eq. (\ref{Eq:c-gamma}), the second line is from the orthonormality of the eigenstates, $\langle \gamma_m^\dagger\gamma_{m'} \rangle = \delta_{mm'}\langle n_m \rangle = \delta_{mm'}f(\varepsilon_m-\mu)$. When a lattice site consists of three components, for example, $l=(x,y,z)$ and $l=(x',y',z')$, by setting $i=(x,y)=(x',y')$ with $z=t$ and $z'=b$, we can obtain the order parameter of $\Delta(i)$ in Eq. (4) at an equilibrium.
The calculated order parameter is fed back to the intersurface interaction Hamiltonian of Eq. (3), and, as a result, the order parameter is obtained self-consistently. 
As we point out in the text, our Hamiltonian has both orbital and spin degrees of freedom and there are four possible pairings in the order parameter described in Eq. (4). In order to clarify this point, we define exciton order parameter subset as $\mathbf{\Delta}_{\tau,\sigma},~\mathbf{\Delta}_{\tau',\sigma},~\mathbf{\Delta}_{\tau,\sigma'}$, and $\mathbf{\Delta}_{\tau',\sigma'}$ where $\tau~(\sigma)$ stands for the same orbital (spin) pairing, while $\tau'~(\sigma')$ stands for different orbital (spin) pairing (e.g. $\mathbf{\Delta}_{\tau,\sigma}\supset\lbrace \Delta_{A\uparrow,A\uparrow}~\Delta_{B\uparrow,B\uparrow}~\Delta_{A\downarrow,A\downarrow}~\Delta_{B\downarrow,B\downarrow}\rbrace$).
%%%%%%%%%%%%%%%%%%%%%%%%%%%%%%%%%%%%%%%%%%%%%%%

\section{Ground State Energy Calculation} \label{A:Etot}
\begin{figure}[t!] %FIGURE1A=============================
\centering
\includegraphics[width=0.5\textwidth]{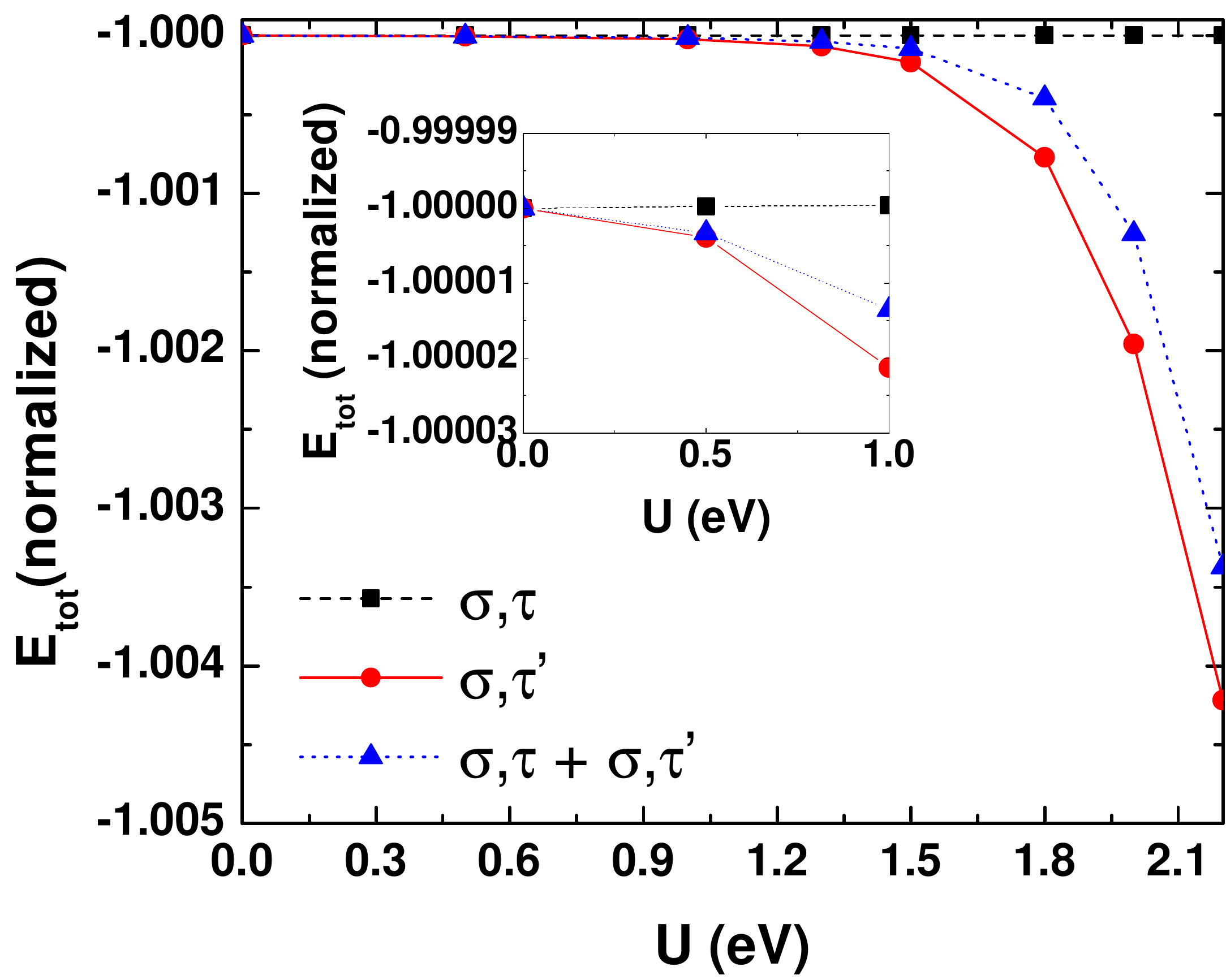}
\caption{The total energy of the system as a function of interaction constant $U$. The ground state energy is calculated self-consistently at equilibrium and normalized by $\vert E_{tot} \vert$ at $U=0~eV$. Inset magnifies part of the plot.}\label{figA_Etot}
\end{figure}

In the mean-field approach, the intersurface pairing scheme whose ground state energy is the lowest will be the energetically favorable pairing term. We perform numerical calculations and immediately find that the pairing $\mathbf{\Delta}_{\tau,\sigma'}$ and $\mathbf{\Delta}_{\tau',\sigma'}$ are zero. In order to determine which pairing term provides the lowest ground state among $\mathbf{\Delta}_{\tau,\sigma}$ and $\mathbf{\Delta}_{\tau',\sigma}$, we perform numerical calculations and obtain the total energy via, 
\begin{equation} \label{Eq:Etot}
E_{tot}=\langle\mathbf{H} \rangle = \sum_\alpha \varepsilon_\alpha - U\sum_{i,s,s'}\vert\Delta_{s,s'}(i)\vert^2,
\end{equation}
where $\epsilon_\alpha$ is an eigenvalue of the total Hamiltonian obtained from Eq. (\ref{Eq:VDV}). The index $\alpha$ runs over the occupied states and $\Delta_{s,s'}(i)$ is defined in Eq. (4). The ground state energy in equilibrium is self-consistently calculated and the result is presented in Fig. \ref{figA_Etot}. The lowest energy is obtained when we choose $\mathbf{\Delta}_{\tau',\sigma}$, which is the intersurface pairing between the same spins but different orbitals.
 %FIGURE1A=================================
%%%%%%%%%%%%%%%%%%%%%%%%%%%%%%%%%%%%%%%%%%%%%%%

\section{Estimation of the Intersurface Interaction $U$} \label{A:U}
The bare Coulomb intersurface interaction strength with a consideration of dielectric environment in inset of Fig. 2(b) is described by Eq. (6). In order to generalize the argument, we neglect finite-size effect and assume $x$ and $y$ as a periodic. Consequently, the real space expression is obtained from Fourier transform analysis in continuum limit as
\begin{equation}\label{Eq:Vtb2}
\begin{aligned}
U_{tb}(r)&=\frac{1}{(2\pi)^2}\int d^2\mathbf{q}~ \tilde{U}_{tb}(q)e^{i\vec{q}\cdot\vec{r}} \\
&=\frac{1}{2\pi}\int_0^{\infty} dq \frac{8\pi e^2}{D(q)}\epsilon_{TI} \sum_{l=0}^\infty \frac{(-1)^l}{2^{2l}l!~l!}(qr)^{2l}. \\
\end{aligned}
\end{equation}
From the first to second line of the Eq. (\ref{Eq:Vtb2}), we evaluate radial part integration by using the Bessel function \cite{Abramowitz1972}, ${\rm J}_0(z)=1/(2\pi)\int_0^{2\pi}d\theta~e^{iz\cos\theta}=\sum_{l=0}^\infty \frac{(-1)^l}{2^{2l}l!~l!}(z)^{2l}$. As we are interested in the on-site intersurface interaction only, the in-plane radius is set to be $r=0$ and corresponding real space expression of the Coulomb interaction in Eq. (2) is 
\begin{equation}\label{Eq:Vtb3}
U=U_{tb}(0)=\frac{1}{2\pi}\int_0^{\infty} dq \frac{8\pi e^2}{D(q)}\epsilon_{TI}.
\end{equation}
We use the material parameters of $\epsilon_{G1}=\epsilon_{G1}=3.9\epsilon_0$ (SiO$_2$), where $\epsilon_0$ is a vacuum dielectric constant. Assuming a linear dispersion relationship at the surfaces, the Fermi wavevector is calculated as $k_F=E_F/\hbar v_F$, where $\hbar v_F=3~eV\AA$ and $E_F=V_g=1.0~eV$ (potential induced by a gate bias at each surfaces). By setting $\epsilon_{TI}$ as a variational parameter with fixed thickness of $d=24~\AA$, the numerical integration is performed and the result is shown in Fig. 2(b).

\begin{figure}[t!] %FIGURE1A=============================
\centering
\includegraphics[width=0.5\textwidth]{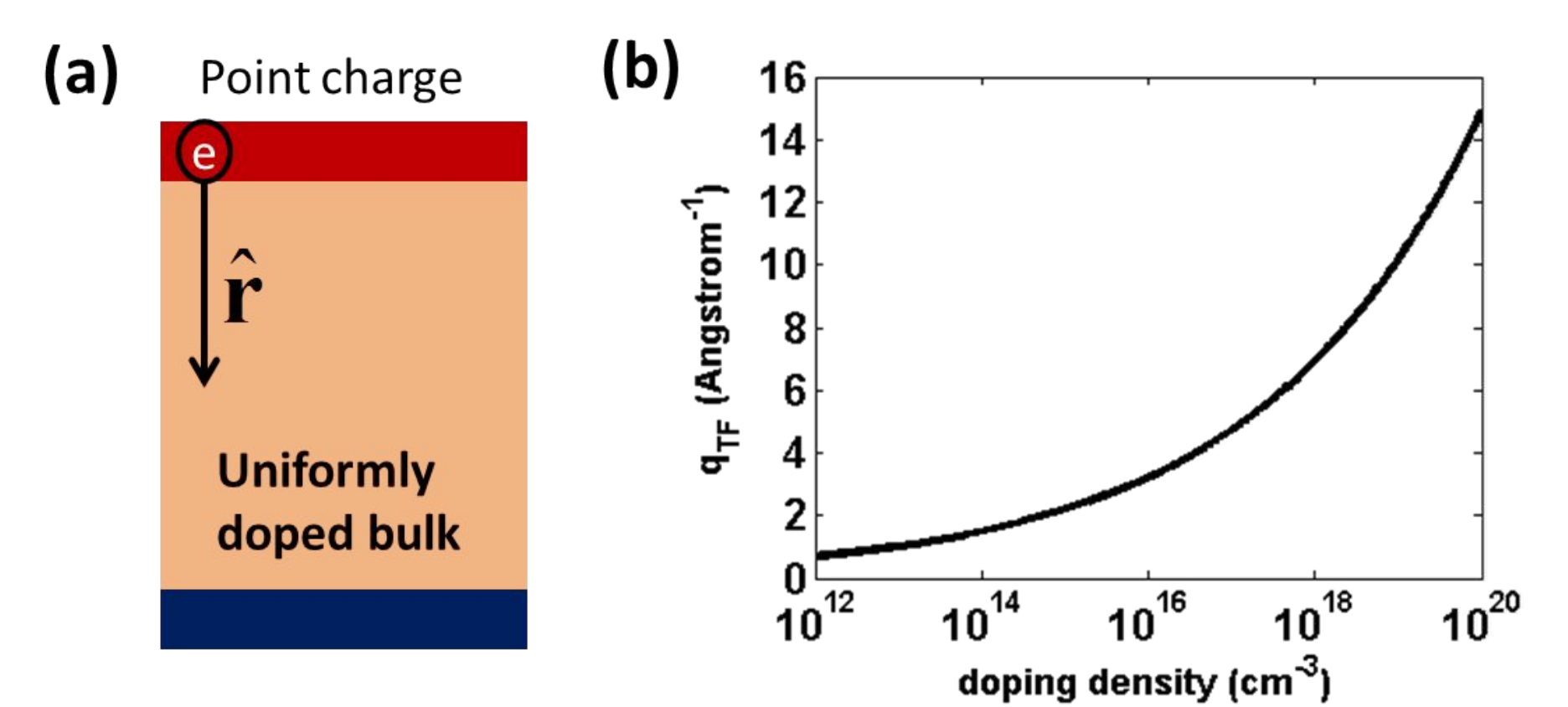}
\caption{(a): Point charge potential at the surface is screened by constant doping in the bulk. (b): The resultant $q_{TF}$ as a function of doping level in bulk. $x$-axis is plotted in a logarithm scale.}\label{figA_TF}
\end{figure}

In addition, we estimate a screening effect of the bulk doping on the intersurface interaction. By simplifying the problem as a point charge like particle screened by the constant background doping as illustrated in Fig. \ref{figA_TF}(a), we calculate Thomas-Fermi wavevector\cite{Ashcroft1976}:
\begin{equation}
q_{TF}=\frac{2.95}{(r_s/a_0)^(1/2)} ~\AA^{-1},
\end{equation}
where the free electron sphere, $r_s$, and the effective Bohr radius, $a_0$, are
\begin{equation}\label{Eq:ra}
r_s=\left( \frac{3}{4\pi n}\right)^{1/3},~a_0=\frac{4\pi \epsilon_{TI} \hbar^2}{m^* e^2}.
\end{equation}
In Eq. (\ref{Eq:ra}), $n$ is electron density and $m^*$ is an effective mass. Using the effective mass\cite{SpringerBiSe} of Bi$_2$Se$_3$ as $m^*\simeq0.155 m_e$ and dielectric constant of $\epsilon_{TI}=100\epsilon_0$, the resultant $q_{TF}$ with various doping level is presented in Fig. \ref{figA_TF}(b). The bulk doping effectively screens the intersurface interaction as the doping level increases. 
 %FIGURE1A=================================
%%%%%%%%%%%%%%%%%%%%%%%%%%%%%%%%%%%%%%%%%%%%%%%

\section{Terminal and Spatially Resolved Current Calculation} \label{A:I}
In case of the channel connected to the contact $1$ and $2$, we calculate the current by the multi-channel Landauer-B\"{u}ttiker formula in the limit of coherent transport:
\begin{equation}
I(V_{12})=\frac{2e}{h}\int{T_{12}(E)[f_1(E)-f_2(E)]}.
\end{equation}
where $V_{12}$ is the potential difference between two contact, $T_{12}$ is the transmission and $f_{1(2)}$ is Fermi-Dirac distribution of the contact $1$ ($2$). In the non-equilibrium Green's function (NEGF) formalism, it is also possible to calculate the spatially resolved current from point $\mathbf{r}_1$ to $\mathbf{r}_2$ is evaluated by using\cite{Datta:2000, Datta2005}
\begin{equation} \label{Eq:Ix}
\begin{aligned}
&I(\mathbf{r}_1\rightarrow \mathbf{r}_2)=\frac{ie}{\hbar} \int \frac{dE}{2\pi}  \\
&~~~~~~~\left[H(\mathbf{r}_{12}) ( G^n(\mathbf{r}_{21},E) - G^p(\mathbf{r}_{21},E) ) \right. \\
&~~~~~~~\left. -H(\mathbf{r}_{21}) ( G^n(\mathbf{r}_{12},E) - G^p(\mathbf{r}_{12},E) ) \right]. \\
\end{aligned}
\end{equation}
In Eq. (\ref{Eq:Ix}), $G^{n(p)}$ is the electron (hole) correlation functions calculated with NEGF method, and $G(\mathbf{r}_{12})$ and $H(\mathbf{r}_{12})$ represents the off-diagonal block connecting sites $\mathbf{r}_1$ and $\mathbf{r}_{2}$ which is only nonzero for nearest neighbors.  

\bibliographystyle{apsrev}
\bibliography{TI_sp_cohr_references}

\end{document}